\documentclass[prl,aps,amsmath,amssymb,amsfonts,floatfix,showpacs,twocolumn]{revtex4-1} 
\usepackage{graphicx}
\usepackage{amsmath,amsfonts}
\usepackage{xcolor}
\usepackage{color}
\usepackage{mathtools}
\usepackage{eufrak}
\usepackage{pgfplots}
\usepackage{multirow}
\usepackage{textcomp}
\usepackage{units} 
\usepackage{xfrac}

\newcommand{\RN}[1]{\textup{\uppercase\expandafter{\romannumeral#1}}}

\def\presuper#1#2%
  {\mathop{}%
   \mathopen{\vphantom{#2}}^{#1}%
   \kern-0.5\scriptspace%
   #2}

\usepackage{hyperref}
\hypersetup{colorlinks=true,breaklinks,linkcolor=blue,urlcolor=blue,citecolor=blue}
\begin{document}

\author{M.~Reitner$^a$, P.~Chalupa$^a$, L.~Del~Re$^{b,c}$, D.~Springer$^{a,d}$, S.~Ciuchi$^e$, G.~Sangiovanni$^f$, and A.~Toschi$^a$}

\affiliation{$^a$Institute of Solid State Physics, TU Wien, 1040 Vienna, Austria}
\affiliation{$^b$ Department of Physics, Georgetown University, 37th and O Sts., NW, Washington, DC 20057, USA}
\affiliation{$^c$  Erwin Schr\"{o}dinger International Insitute for Mathematics and Physics, Boltzmanngasse 9 1090 Vienna, Austria}
\affiliation{$^d$ Institute of Advanced Research in Artificial Intelligence, IARAI, A-1030 Vienna, Austria}
\affiliation{$^e$ Dipartimento di Scienze Fisiche e Chimiche, Universit\`a dell'Aquila, and Istituto dei Sistemi Complessi,CNR, Coppito-L'Aquila, Italy}
\affiliation{$^f$  Institut f\"ur Theoretische Physik und Astrophysik and W\"urzburg-Dresden Cluster of Excellence ct.qmat, Universit\"at W\"urzburg, 97074, Germany}

\title{Attractive effect of a strong electronic repulsion -- the physics of vertex divergences}

\begin{abstract}
While the breakdown of the perturbation expansion for the many-electron problem has several formal consequences, here we unveil its physical effect: Flipping the sign of the effective electronic interaction in specific scattering channels. By decomposing local and uniform susceptibilities of the Hubbard model via their spectral representations, we prove how entering the non-perturbative regime causes an enhancement of the charge response, ultimately responsible for the phase-separation instabilities close to the Mott MIT. Our analysis opens a new route for understanding phase-transitions in the non-perturbative regime and clarifies why attractive effects emerging from a strong repulsion can induce phase-separations, but not $s$-wave pairing or charge-density wave instabilities.
\end{abstract}

\maketitle


{\sl Introduction} --  
While the many-electron problem of condensed matter and QED are similar in several respects (e.g., their
Feynman diagrammatic description), they differ in a very important point: 
For the former, no small expansion parameter can be identified a priori. 

The applicability of weak-coupling approaches in condensed matter 
depends, in fact, on how efficiently the Coulomb interaction is screened in the compounds under consideration.
This often requires to go beyond the comfort zone of a perturbative description with important formal and algorithmic implications, intensely discussed in the recent literature\cite{Schaefer2013,Janis2014,Kozik2015,Stan2015,
Ribic2016,Gunnarsson2016,Schaefer2016c,Gunnarsson2017,Tarantino2018,Vucicevic2018,Rohringer2018,Chalupa2018,Thunstroem2018,
Kugler2018,Krien2019,Krien2019SBE,Springer2020,Chalupa2020,Schaefer2020}.

\begin{figure*}[th!]
  \centering
  {{\resizebox{9.5cm}{!}{\includegraphics {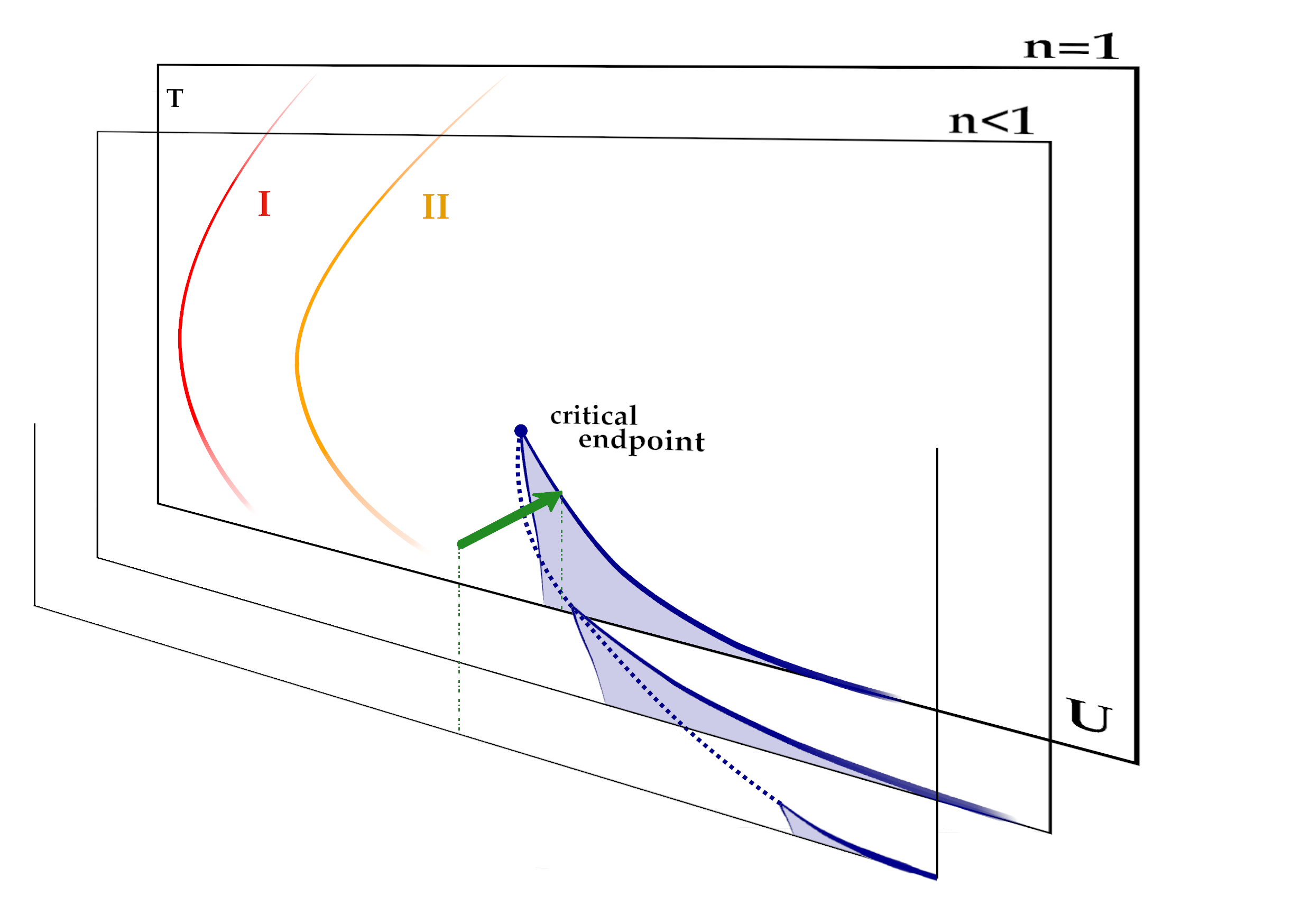}}}} \hspace{-10mm}
  {{\resizebox{7.9cm}{!}{\includegraphics {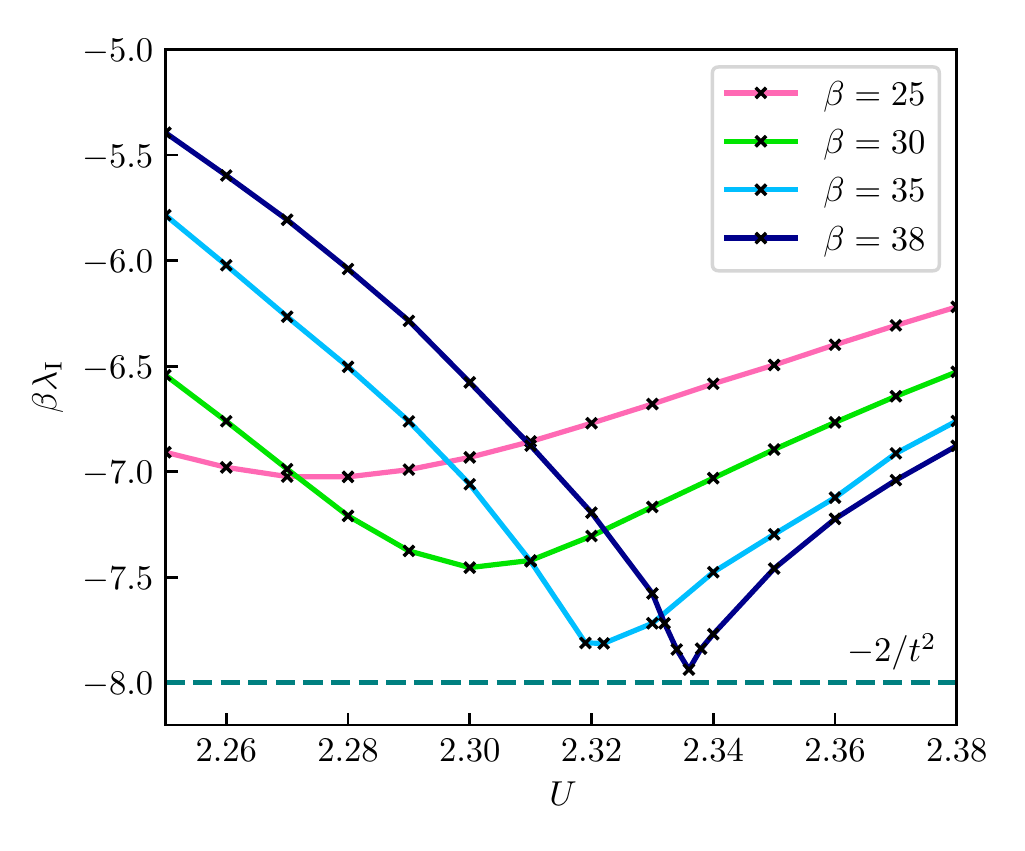}}}} \vspace{-6mm}
  \caption{{\small Left: sketchy representation of the Hubbard model phase-diagram in DMFT (coexistence region of the Mott MIT at $n=1$/ phase-separation at $n<1$: shadowed blue regions; corresponding critical endpoints: blue dot/ dotted blue line; first(I)/second(II) line of divergences of $\Gamma^{\nu\nu'}$ at half-filling: red/orange curve). Right:  lowest eigenvalue of  $\chi^{\nu \nu'}$ for different temperatures, computed by DMFT on a half-filled Bethe-lattice (solid lines), compared with the divergence condition $-\frac{2}{t^2}$ (dashed line) of the analytical expression for the susceptibility [Eq.~(\ref{eq:kappaDMFT})], see text.}}
  \vspace{-2mm}
\label{Fig:1}
  \end{figure*}
  
In this paper, we demonstrate that the breakdown of perturbation theory\cite{Schaefer2013,Kozik2015,Gunnarsson2017} 
is not a mere formal issue, but
that it is directly linked to precise physical effects of high importance for correlated electron systems. 
In particular, we will show {\sl how}, and {\sl to what extent}, entering the non-perturbative regime can turn a strong electrostatic repulsion into an effective attraction. 

{\sl  Non perturbative regime} --  To go beyond the weak-coupling framework, we exploit one of the most successful many-body methods, which does not rely on perturbation expansion: We will consider the dynamical mean-field theory (DMFT)\cite{Georges1996} applied to the Hubbard model. 
In particular, we will focus on the charge response, as this directly reflects the action of a density-density interaction and its screening.
 Its local (and static) part is defined\cite{Bickersbook2004,Rohringer2012,SM}
 as $\chi_{\rm loc} =\int_0^\beta  \, d\tau [  \langle \hat{n}(\tau) \hat{n}(0) \rangle - \langle \hat{n} \rangle^2 ]$ (with $\beta \! = \! (k_BT)^{-1}$), which can be computed by summing 
the corresponding generalized two-particle susceptibility $\chi^{\nu \nu'} (\Omega \!=\! 0)$  (at zero transfer frequency $\Omega$) over all the fermionic Matsubara frequencies $\nu, \nu'$:
 \begin{eqnarray} 
\chi_{\rm loc} & = & \frac{1}{\beta^2} \sum_{\nu \nu'} \chi^{\nu \nu'}\!(\Omega \!= \! 0) \\
  & = & \sum_{\alpha} \lambda_\alpha w_\alpha. 
\label{eq:chiloc}
\end{eqnarray}
The sum in the second line is recasted\cite{Gunnarsson2017,Springer2020} in the eigenbasis of  $\chi^{\nu \nu'}\!(\Omega\!=\!0)$ with eigenvalues $\lambda_\alpha$, and spectral weights $w_\alpha$, defined through the eigenvectors as $w_\alpha \! = \left[\sum_\nu V^{-1}_\alpha(\nu) \right] \left[\sum_{\nu'} V_\alpha(\nu') \right]$. 

We start from the easiest situation of a half-filled, particle-hole symmetric model, where $\chi^{\nu \nu'}(\Omega\!=\!0)$  is a real, bisymmetric matrix with real $\lambda_\alpha$ and $w_\alpha \ge 0$\cite{Springer2020}.
In this case, it was already shown that the progressive suppression of local  charge fluctuations (i.e., of $\chi_{\rm loc}$) by increasing $U$ is driven by a corresponding decrease of  the eigenvalues $\lambda_\alpha$\cite{Gunnarsson2017,Springer2020}. 
In fact, while the 
$\lambda_\alpha$ are all positive for $U\!=\!0$,  some of them cross zero upon increasing $U$, becoming negative and, hence, strongly reducing the overall value of $\chi_{\rm loc}$.  Each sign-change of one of the $\lambda_\alpha$ corresponds -per definition- to a divergence of the irreducible vertex $\Gamma^{\nu \nu'} = [\chi^{\nu \nu'}]^{-1} - [\chi_0^{\nu\nu'}]^{-1}$ or equivalently, to a non-invertibility of the associated Bethe-Salpeter equation (BSE)\cite{Schaefer2013,Schaefer2016c}. 
Exactly for the same parameter sets, one also observes a crossing of solutions in the Luttinger-Ward 
functional\cite{Kozik2015,Gunnarsson2017}. 
The parameters, where the {\sl lowest} $\lambda_\alpha$ crosses zero, thus mark\cite{Gunnarsson2017} 
the end of the perturbative regime.
We recall that a similar fate occurs to the local pairing fluctuations and, hence, to the BSE in the particle-particle channel\cite{Schaefer2013,Schaefer2016c,Springer2020}.

While the important, but technical question of how cutting-edge algorithms (especially those based on irreducible vertices\cite{Toschi2007,Ayral2016} or bold resummations of Feynman diagrams\cite{Kozik2015,Vucicevic2018}) get affected by this has been the subject of many recent studies, from a more physical point of view it seems natural\cite{Gunnarsson2017} to relate such non-perturbative manifestations to a suppression of the corresponding local fluctuations. 
However, a different viewpoint is possible\cite{Reza2019}: as the irreducible vertex $\Gamma$ is the core of a BSE, its multiple sign-changes 
(driven by those of $\lambda_\alpha$\cite{Chalupa2018}) could be interpreted as a flipping of a repulsive into an attractive interaction (or vice versa).
Heuristically, if we consider a simple RPA-like  expression  [$\Gamma^{\nu \nu'} \! \rightarrow \! \Gamma_0 > 0$ (const.)] for the charge and the pairing fluctuations, $\chi_{\mathbf{q}}=\chi^0_{\mathbf{q}}(1+\Gamma_0 \chi^0_{\mathbf{q}})^{-1}$,
a sign-change of $\Gamma_0$ would induce an enhancement, instead of a suppression, of the corresponding susceptibility with increasing interaction.

Though intriguing, this interpretation raises additional questions: it seems hard to be reconciled with suppression of fluctuations at half-filling discussed above and it may lead to rather bizarre physical predictions, if carried to its extreme consequences. For instance, one would expect the multiple divergences of $\Gamma$ found\cite{Schaefer2016c} in the phase-diagram of the Hubbard model to be reflected in a series of maxima  (maybe even of divergences) of the charge and pairing susceptibilities by increasing $U$.
However, such a peculiar oscillatory behavior has never been reported\cite{Kotliar2002,Werner2007,Eckstein2007,Sordi2011,
Sordi2019,Reza2019,Walsh2019}.

{\sl A glimpse from DMFT} -- As known\cite{Kotliar2002,Werner2007,Eckstein2007,Reza2019}, DMFT calculations show that charge fluctuations are strongly enhanced in the proximity of the critical endpoint of the Mott metal-insulator transition (MIT) of the Hubbard model\cite{SM} (blue dot, topping the shadow area in the $n=1$ plane of the phase-diagram sketched in Fig.\,\ref{Fig:1}). Specifically, while at half-filling the isothermal compressibility $\kappa$  decreases monotonically with increasing $U$,  a strongly enhanced compressibility is observed in the crossover region at {\sl finite} doping on both sides of the MIT. In fact, $\kappa$ even diverges along two curves in the parameter space embracing the critical endpoint of the MIT (blue dotted in the left panel of Fig.~\ref{Fig:1}, shown on one side only) and marking the onset of a  phase separation at lower $T$. 

As we will show, this behavior of $\kappa$ is directly linked to the divergences of the irreducible vertex $\Gamma$ 
and, specifically, to the first ones encountered\cite{Schaefer2013,Schaefer2016c, Springer2020} in the correlated metallic region, much before the MIT itself. The location is sketched as red (I) and orange (II) curves in the $n=1$ plane of Fig.~\ref{Fig:1}.

In general, the compressibility $\kappa$ can be defined (i) at the one-particle level,  as  the derivative of the density  w.r.t. the chemical potential ($\frac{d n}{d \mu}$)
or  (ii) at the two-particle level, as the static limit (${\bf q } \! \rightarrow \! 0, \Omega=0$) of the momentum/frequency dependent charge response function  $\chi_{\bf q} (\Omega)$, obtained through the BSE\cite{Georges1996}
\begin{equation}
\chi_{\bf q}(\Omega) = \frac{1}{\beta^2} \sum_{\nu \nu'} \left[ [\chi^{0}_{\bf q}(\Omega)]_{\nu\nu'}^{-1} + \Gamma^{\nu \nu'}\!(\Omega)]\right]^{-1},
\label{eq:BSEDMFT}
\end{equation}
where the bubble term reads $\chi^{0,\nu\nu'}_{\bf q}(\Omega)\!=\!-2 \beta \sum_{\bf k}  G({\bf k}, \nu) G({\bf k}\!+\!{\bf q}, \nu\!+\!\Omega) \delta_{\nu\nu'}$.
In DMFT, where the self-energy and the irreducible vertex $\Gamma$ are both extracted from a (self-consistently determined) auxiliary impurity model\cite{Georges1996}, 
the two definitions yield  per construction the same value of $\kappa$ (see Ref.~\cite{vanLoon2015} and \cite{Krien2017,Reza2019,Krien2019}).

\begin{figure}[t!]
\centering
{{\resizebox{8.5cm}{!}{\includegraphics {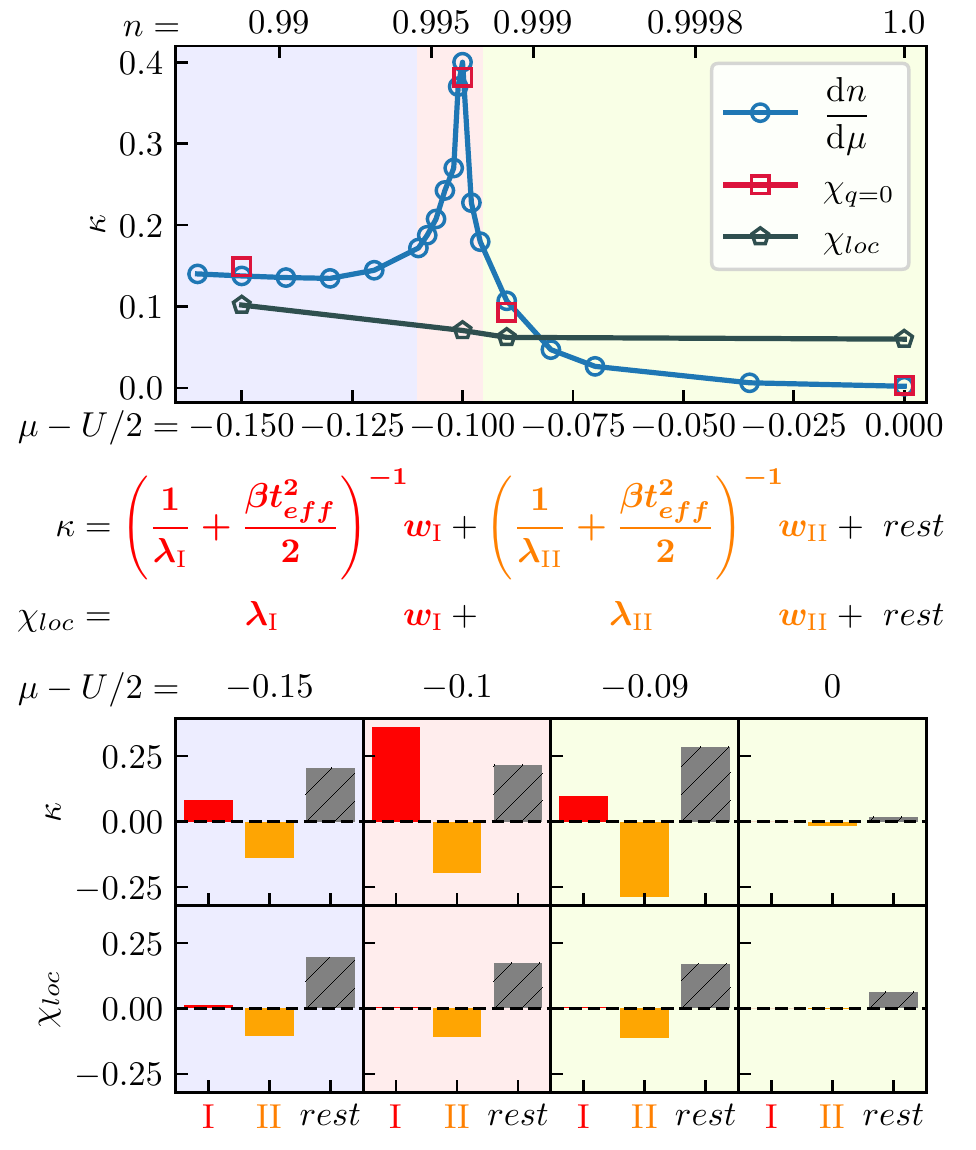}}}}
\caption{{\small {\sl Top}: compressibility $\kappa$ (blue circles from the numerical derivative of $n$ w.r.t.~$\mu$; red squares from the BSE for $\chi_{{\bf q}=0}$)  and local charge susceptibility ($\chi_{loc}$) of the Hubbard model computed in DMFT for $\beta\!=\!53$ and $U\!=\!2.4$ (on a square lattice with half-bandwidth $D=1$). {\sl Bottom}: 
analysis of the contributions to $\kappa$ and $\chi_{\rm loc}$ arising from the lowest two {\sl real} eigenvalues (``I" in red, ``II" in orange) and from all the remaining terms (``$rest$" in grey) for four different dopings. The light background colors are just a guide to the eye.}}
\label{fig:2}
\end{figure}

The locality of $\Gamma$ in Eq.~(\ref{eq:BSEDMFT}) makes the relation between local and collective properties particularly transparent in DMFT. 
In fact, by straightforwardly extending\cite{SM} a famous result of Ref.~\cite{Georges1996} for the charge channel, we obtain the following analytical expression:
 \begin{equation}
 \kappa=  \sum_{\alpha} \left(\frac{1}{\lambda_{\alpha}} +   \nicefrac{1}{2} \beta t^2\! \right)^{-1}\!  \! w_\alpha \, ,
 \label{eq:kappaDMFT}
 \end{equation}
 which holds exactly for the Bethe-lattice case (here of half-bandwidth $D=2t=1$), 
 independently of its filling.  As we will discuss below, it also represents a very good approximation if the DMFT is performed on other, more realistic lattices\cite{SM}.
A quick glance at Eq.~(\ref{eq:kappaDMFT}) immediately shows that the only possibility for a divergence of $\kappa$ is that the condition $\beta \lambda_{\alpha} = - \frac{2}{t^2}  \! < \! 0 $ is verified for one eigenvalue of $\chi^{\nu \nu'}\!(\Omega \!= \!0)$. Evidently, this locates necessarily such divergences of $\kappa$ on the right side of the first vertex-singularity line (red curve in Fig.\,\ref{Fig:1}) and defines precise constraints, calling for a quantitative analysis.

\begin{figure}[th!]
  \centering
  {{\resizebox{7.cm}{!}{\includegraphics {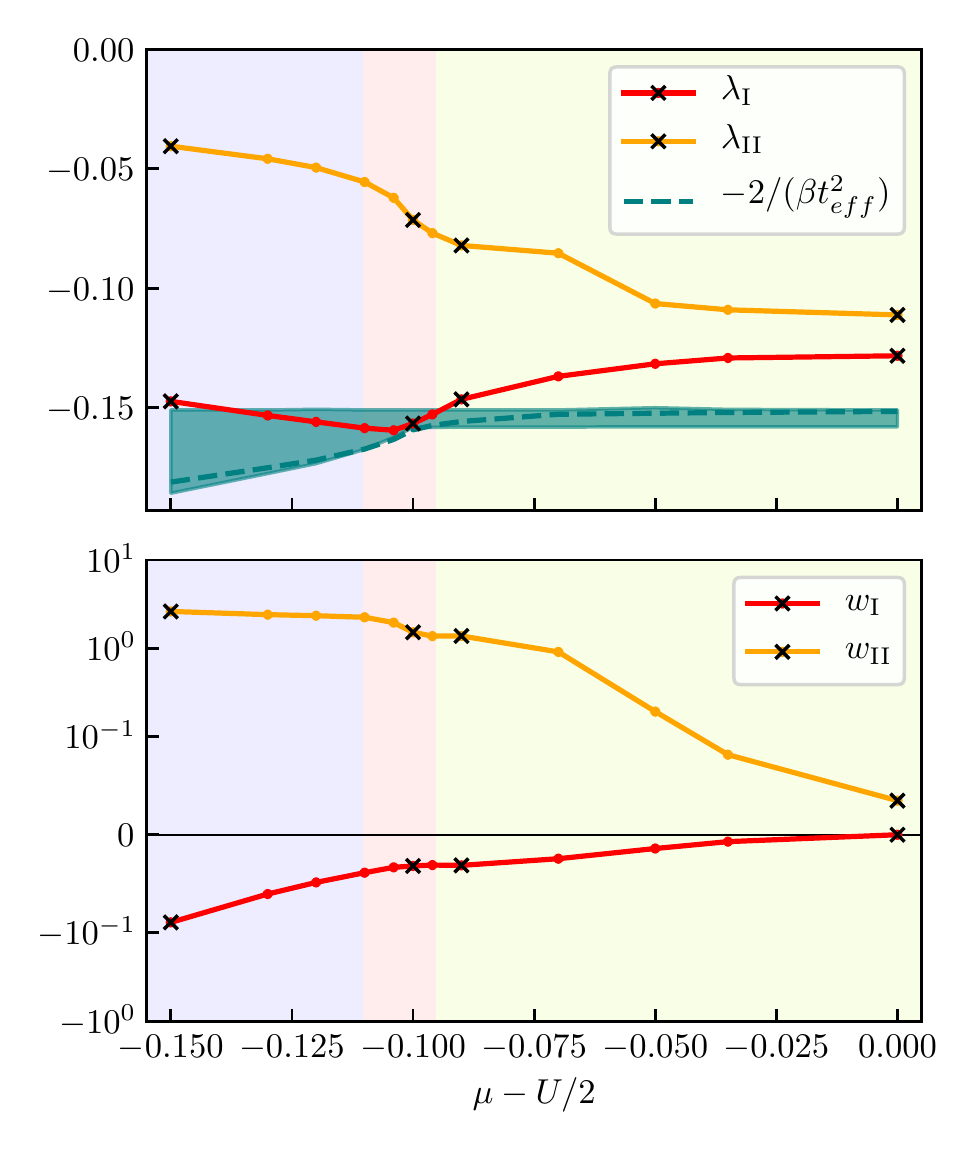}}}  \vspace{-5mm}
  \caption{{\small Top: lowest two real eigenvalues ($\lambda_{\rm I}$ in red, $\lambda_{\rm II}$ in orange) of the local charge susceptibility for the same parameters as Fig.~\ref{fig:2}; the weak frequency dependence of the bubble term for the square lattice case is marked, for each $\mu$, by a blue-shaded area; 
  the values of $\nicefrac{-2}{\beta t_{\rm eff}^2}$\cite{SM} by a dashed line. 
   Bottom: corresponding spectral weights ($w_{\rm I}$ and $w_{\rm II}$) on logarithmic y-axis. }}
\label{fig:3}
}
\end{figure}

{\sl The half-filling case} --  We consider first the (particle-hole symmetric) half-filled Bethe-lattice case, computing the evolution of the lowest eigenvalue $\lambda_{\rm I}$ as a function of $U$ for different temperatures (right panel of 
Fig.\,\ref{Fig:1}). 
In particular, we use a continuous-time quantum Monte Carlo (CT-QMC) solver provided by the w2dynamics package\cite{w2dynamics} throughout this work to obtain the one- and two-particle quantities (for details, see [\onlinecite{SM}]).
As discussed in the literature\cite{Chalupa2018,Springer2020}, due to the high-symmetry of this case, $\lambda_{\rm I}$ is associated to a real,  antisymmetric eigenvector ($V_{\rm I}(\nu) = - V_{\rm I}(-\nu)$, hence $w_{\rm I}=0$). 
From the data of Fig.~\ref{Fig:1}b, we clearly see that  $\lambda_{\rm I}$ displays a minimum at intermediate $U$, in the {\sl crossover region} of the Mott MIT. By reducing $T$ the minimum gets sharper and progressively closer to the necessary condition of a divergence of $\kappa$ (marked by dashed line).  Remarkably, the condition is fulfilled at the (second-order) critical endpoint of the MIT (at $U \! \simeq \! 2.33$, $\beta  \! \simeq \! 38$), where  the minimum of $\lambda_{\rm I}$ becomes a cusp, before one starts observing a coexistence of two solutions at lower $T$. We note that this behavior can alternatively be understood from the critical properties of the MIT, as independently proven by van Loon and Krien\cite{vanLoon2020}.
At half-filling, however, the divergence of $(1/\lambda_{\rm I} + \nicefrac{1}{2} \beta  t^2)^{-1}$ does not have {\sl any} physical effect on $\kappa$, because the associated spectral weight $w_{\rm I}$ in Eq.~(\ref{eq:kappaDMFT}) is always zero,  due to the perfect antisymmetry of $V_{\rm I}(\nu)$.
We note that the second lowest eigenvalue ($\lambda_{\rm II}$), associated with a symmetric eigenvector [$V_{\rm II}(\nu) \! = \! V_{\rm II}(-\nu)$] 
becomes also negative (after the orange curve in Fig.\,\ref{Fig:1}), 
but it never reaches the critical condition $\beta \lambda_{\rm II} \! =  \! - \frac{2 \; \,}{t^2}$. In fact, as its 
spectral weight is positive, it contributes to a progressive suppression of $\kappa$.

 \begin{figure*}[th!]
  \centering
  {{\resizebox{17cm}{!}{\includegraphics {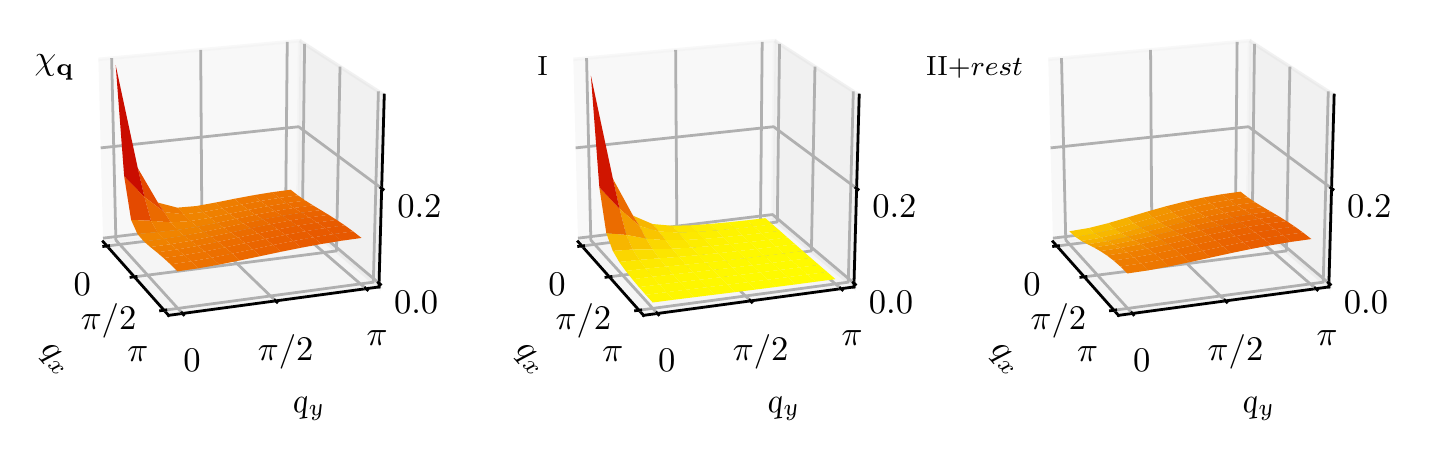}}}  \vspace{-5mm}
  \caption{{\small Left: Momentum-dependence of the charge susceptibility $\chi_{{\bf q}}$ computed in DMFT on a square lattice for $U=2.4$, $\beta= 53$ for $\mu-U/2 = -0.1$, corresponding to the maximum of the compressibility $\kappa$. Center: 
contribution stemming from $\lambda_{\rm I}$. Right: all other contributions summed.}}
\label{fig:4}
}
\end{figure*} 

{\sl Out of half-filling} -- The results above crucially depend on the high-symmetry properties\cite{DelRe2019,Springer2020} 
of the (non-frustrated) half-filled case. 
As soon as those are lifted, e.g. by doping the system and/or adding a next-to-nearest neighbor
hopping term ($t'$) striking changes are observed.
We consider explicitly the case of a hole doped system ($\mu \! - \! \frac{U}{2} \! < \!0$, $n \! < \! 1, t' \! = \!0$) on a square lattice (with half-bandwidth $D=1$)
in the {\sl crossover} region of the phase-separation near the critical endpoint of the half-filling MIT (i.e., $U \!= \!2.4$, $\beta \! =  \!53$, as schematically indicated by the green arrow in Fig.\,\ref{Fig:1}). 
In Fig.~\ref{fig:2}, we report the behavior of the local ($\chi_{\rm loc}$) and the uniform ($\kappa$) charge susceptibility as a function of the chemical potential,
varying it towards half-filling ($\mu=\frac{U}{2}$ on the right side). 
 Our data show a clear {\sl dichotomy} in the behavior of $\chi_{\rm loc}$ and $\kappa$. While  $\chi_{\rm loc}$, directly evaluated from Eq.~(\ref{eq:chiloc}), 
gets monotonically suppressed towards half-filling, where correlations are stronger,  $\kappa$, evaluated both as numerical derivative as well as from Eq.~(\ref{eq:BSEDMFT}),  displays a prominent maximum at a finite doping: 
This indicates that we are in the crossover region, just slightly above the critical endpoint of the phase-separation (dotted line in the sketch of Fig.~\ref{Fig:1}). 
  
{\sl Diagnostics of $\kappa$} -- A clear-cut theoretical insight into this phenomenology is obtained by decomposing $\chi_{\rm loc}$ and $\kappa$, computed at several dopings, in terms of the contributions stemming from the different eigenvalues $\lambda_{\alpha}$ of $\chi^{\nu\nu'}$, in the spirit of Refs.\cite{Gunnarsson2015,Valli2015,Gunnarsson2016,Wu2017,Stepanov2019,Kauch2020,Rohringer2020}. Here, this procedure, which is always possible numerically, allows for a very transparent analytical understanding, based on the Bethe-lattice expression, Eq.~(\ref{eq:kappaDMFT}). In fact, the deviations found for the square lattice case are marginal in the parameter region of our interest\cite{SM}. Eq.~(\ref{eq:kappaDMFT}) can be thus exploited, in an approximated form ($t^2  \rightarrow t^2_{\rm eff}$, where $t_{\rm eff}$ weakly depends on $\mu$\cite{SM}), as a key to the interpretation.
 
We start by separating  $\chi_{\rm loc}$  in terms of the two lowest real $\lambda_\alpha$-contributions and the rest to the sum in Eq.~(\ref{eq:chiloc}). 
As shown in the bottom panel of Fig.\,\ref{fig:2} at finite doping one observes a tiny {\sl positive} contribution from $\lambda_{\rm I}$  (red bar) enhancing $\chi_{\rm loc}$, which fully disappears at half-filling where its weight $w_{\rm I} \! =0 \! $ due to symmetry.
The corresponding decomposition for $\kappa$ shows, instead, that precisely the contribution originated from $\lambda_{\rm I}$ is 
responsible for its non-monotonous behavior as well as for the sharp maximum. By comparing the two decompositions, one 
immediately notes how the dichotomy of the local and the uniform charge response is essentially controlled by the contributions 
(red bars) associated to the lowest real eigenvalue of $\chi^{\nu\nu'}$.
 
 The outcome of our analysis can be readily understood in terms of Eq.~(\ref{eq:kappaDMFT}), by studying the behavior of $\lambda_{\rm I}$ and $w_{\rm I}$ for different doping, as reported in Fig.~\ref{fig:3}. 
 If $\lambda_{\rm I}$ becomes negative enough, closely approaching the condition $\beta \lambda_{\rm I} \simeq -\frac{2 \; \;  \;}{t_{\rm eff}^2}$, a maximum of $\kappa$ is observed.
The difference w.r.t.~the half-filled case is that the corresponding weight $w_{\rm I}$ is now {\sl finite}, and actually {\sl negative}, thus contributing to an overall enhancement of the charge response. Because of the small weight $w_{\rm I}$, such an effect is generally mild, unless $\lambda_{\rm I}$ gets negative enough to trigger a strong enhancement or even the divergence of $\kappa$.
Fig.~\ref{fig:3} also shows that the weight associated to the second lowest real eigenvalue ($\lambda_{\rm II}$) always remains positive, as at half-filling. Hence, even if both $\lambda_{\rm I}, \,\lambda_{\rm II}$ are negative, the latter is responsible for a {\sl suppression} of the charge response.
In fact, it is the overall sign of $\lambda_\alpha w_\alpha$, which determines, in general, whether the net effect can be interpreted as repulsive or attractive in the charge sector, since the sign of $w_\alpha$ is no longer positive-definite\cite{SM}.

At the same time, the evolution of $w_\alpha$ of each $\lambda_\alpha$, is smooth in the phase-diagram  (s.~Fig.~\ref{fig:3} and [\onlinecite{SM}]). Hence, crossing the first divergence\cite{Schaefer2013,Kozik2015,Schaefer2016c,Gunnarsson2017,Springer2020} line of $\Gamma^{\nu\nu'}$, which is associated to a sign-change of $\lambda_{\rm I}$, corresponds to flipping the net action of the corresponding contribution ($\lambda_{\rm I} w_{\rm I}$) to the charge response from suppressing to enhancing. 

 We stress that having $w_{\rm I}<0$ is crucial both for the emergence of these strong-coupling phase-instabilities and for the dichotomy between the local and uniform response: The sum in
 Eq.~(\ref{eq:kappaDMFT}) can be recasted as
 \begin{equation}
 \kappa =  \sum_\alpha \frac{\chi_{\rm loc}^\alpha}{1+ \beta J_{\rm eff}^\alpha}
 \label{eq:kapparecast}
 \end{equation} 
 where $\chi_{\rm loc}^\alpha = \lambda_\alpha w_\alpha$ and $J_{\rm eff}^{\alpha} =\frac{t_{\rm eff}^2}{2 w_\alpha }
 \chi^\alpha_{\rm loc}$. 
All summands of Eqs.~(\ref{eq:chiloc})  and (\ref{eq:kapparecast}) are rather similar, except close to the phase-separation where the difference
between the local and uniform response is induced by the first term ($\alpha = {\rm I}$) mainly.
 In that region, as $\chi_{\rm loc}^{\rm I} > 0$,  $w_{\rm I}<0 $ implies a negative coupling
 ($J_{\rm eff}^{\rm I}<0$) in the charge sector.

 {\sl The full momentum dependence} -- We now extend our analysis to the entire momentum dependence of $\chi_{\bf q}$, performed at the same parameter-set where the maximal $\kappa$ is found. In the left panel of Fig.~\ref{fig:4}, where  $\chi_{\bf q}$ is plotted, we observe a rather sharp peak at ${\bf q}\!=\!0$.
In the central and right panels, we decompose $\chi_{\bf q}$ into  the contributions from $\lambda_{\rm I}$  and the remaining eigenvalues, respectively. 
We immediately see that the non-perturbative enhancement of the charge response is confined to the small {\bf q}-sector. Further we note that without the critical, effectively {\sl attractive}, contribution from $\lambda_{\rm I}$, the charge response would have a completely different shape, closely resembling the one at half-filling\cite{SM}: a rather low $\chi_{\bf q}$ with a shallow maximum at ${\bf q}=(\pi,\pi)$. 
This selective enhancement of $\chi_{\bf q}$ around ${\bf q}=0$ increases the corresponding correlation length $\xi$, which is necessary to ensure the second-order nature of the critical endpoints of the phase-separation as well as for inducing the strong dichotomy between the local and the uniform response, discussed above.

We expect the same to happen along the entire, highly non-trivial, path of the phase-separation instability computed in the DMFT phase-diagram of Ref.~\cite{Eckstein2007}.
We also want to stress that the non-perturbative nature associated to the negative sign of $\lambda_{\rm I}$ will prevent all approximations, where the irreducible vertices do not diverge (such as  
RPA, FLEX\cite{Bickersbook2004},  fRG\cite{Metzner2012}, the parquet approximation\cite{Bickersbook2004, Yang2009, Tam2013, 
Valli2015,Li2016, Wentzell2016,Kauch2020}, etc.) to capture this phenomenology.

{\sl Outlook} --  It is insightful to generalize our considerations by extending Eq.~(\ref{eq:kappaDMFT}) to the other sectors, which are mostly reactive to attractive interactions.
One can show\cite{DelRe2020} that the corresponding DMFT expressions for the Bethe-lattice for any static particle-hole susceptibility at ${\bf q} \!= \! {\bf \Pi} = (\pi,\pi,\pi, \ldots)$ (e.g., the CDW in the charge sector), as well as of the pairing ($pp$) s-wave susceptibility at ${\bf q}\!=\!0$ read
\begin{equation}
 \chi_{{\bf q \! = \! \Pi}} = \chi^{pp}_{\bf q \! =\!0} = \sum_{\alpha} \left(\frac{1}{\lambda_{\alpha}} {\bf -} \,   \nicefrac{1}{2} \beta  t^2 \! \right)^{-1} \! w_\alpha,
 \label{eq:otherchis}
 \end{equation}
 independently of the filling.
 This rules out the possibility of inducing CDW or $s$-wave pairing instabilities through a strong local repulsion: divergences of the corresponding responses can only originate from a {\sl large} and {\sl positive} $\lambda_\alpha$, a typical hallmark\cite{Springer2020} of preformed local pairs\footnote{or local moments, if we consider the magnetic sectors}, and hence, of the presence of bare attractive interaction $U \!<\!0$. 
 Here, we clearly see the difference between a bare (and frequency-independent) attractive interaction and an effective one, originating from non-perturbative mechanisms: The effect of the latter can be regarded as truly attractive only in specific sectors and parameter regions.
 
In the future it will be interesting to investigate whether a similar, non-perturbative mechanism is responsible for the enhanced charge fluctuations and phase-separation instabilities reported\cite{DeMedici2017,Chat2020} in extended parameter regions of Hund's metal systems and can, possibly, even trigger the onset of the $s_{\pm}$-pairing.  \\

\begin{acknowledgments}
{\sl Acknowledgments} -- We thank E. van Loon, F. Krien, L. De Medici, M. Capone, T. Sch\"afer, G.~Rohringer, A. Georges and G. Kotliar for insightful discussions. PC, GS and AT also thank the Simons Foundation for the great hospitality at the CCQ of the Flatiron Institute.  The present work was supported by the Austrian Science Fund (FWF) through the project I 2794-N35 and by the U.S. Department of Energy, Office of Science, Basic Energy Sciences, Division of Materials Sciences and Engineering under Grant No. DE-SC0019469. G.S. acknowledges financial support from the DFG through W\"urzburg-Dresden Cluster of Excellence on Complexity and Topology in Quantum Matter - ct.qmat (EXC 2147, project-id 390858490). Calculations have been performed on the Vienna Scientific Cluster (VSC).
P.C. and M.R. contributed equally to this work. \end{acknowledgments}

\bibliography{VERTEX}

{\bf 
\centering{Supplemental Material for:\\ \textit{Attractive effect of a strong electronic repulsion -- the physics of vertex divergences}}} \\

\section{I. analytic understanding of the uniform susceptibility}
In this section we recall the derivation of the momentum and frequency dependent charge response function in DMFT obtained through the Bethe-Salpeter equation (BSE), following Ref.~\cite{Georges1996}. We discuss its analytical properties 
for the Hubbard Model first for the Bethe-lattice and then for the square lattice, by going into the eigenbasis of the generalized local two-particle susceptibility $\frac{1}{\beta^2}\sum_{\nu \nu'}\chi^{\nu \nu'}=\sum_\alpha \lambda_\alpha w_\alpha$, see Eqs.~(1) and (2) of the main text.

The Hamiltonian of the Hubbard model is given by
\begin{equation}
\hat{H} = -\frac{t}{\sqrt{2d}}\sum_{\langle ij \rangle,\sigma}\hat{c}^{\dagger}_{i\sigma}\hat{c}^{\phantom \dagger}_{j\sigma} + U\sum_{i}\hat{n}_{i\uparrow}\hat{n}_{i\downarrow} \, ,
\end{equation}

where $t$ is the hopping between nearest-neighboring sites, U is the local repulsion, and $c_{i\sigma}^{\dagger}/c_{i\sigma}^{\phantom \dagger}$ creates/annihilates an electron with spin $\sigma=\uparrow\downarrow$ on site i with 
$n_{i\sigma}=c_{i\sigma}^{\dagger}c_{i\sigma}^{\phantom \dagger}$.

In the limit of infinite dimensions $d \rightarrow \infty$ the irreducible vertex $\Gamma^{\nu\nu'}(\Omega)$ can be expressed in terms of the local quantities of the auxiliary impurity model
\begin{equation}
\label{equ:gamma}
  \Gamma^{\nu \nu'}(\Omega)=[\chi^{\nu\nu'}\!(\Omega)]^{-1}-[\chi^{\nu\nu'}_0(\Omega)]^{-1}\ ,
\end{equation}
where $\chi^{\nu\nu'}(\Omega)=2(\chi_{\uparrow\uparrow}^{\nu\nu'}(\Omega)+\chi_{\uparrow\downarrow}^{\nu\nu'}(\Omega))$ and 
$\chi_{\sigma\sigma'}^{\nu\nu'}(\Omega)$ is defined as\cite{Rohringer2012}
\begin{eqnarray}
\label{equ:form_gen_chi}
\chi_{\sigma \sigma'}^{\nu \nu'} (\Omega) &=& \int \limits_0^\beta d \tau_1 d\tau_2 d\tau_3 \,
e^{-i\nu \tau_1} 
e^{i(\nu + \Omega)\tau_2} e^{-i(\nu' + \Omega)\tau_3} \nonumber  \\
&\times & [ \langle T_{\tau} c_{\sigma}^{\dagger} (\tau_1)
c_{\sigma}^{\phantom \dagger}(\tau_2) c_{\sigma'}^{\dagger}(\tau_3) c_{\sigma'}^{\phantom \dagger}(0) 
\rangle \\
& -& \langle T_{\tau} c_{\sigma}^{\dagger} (\tau_1)
c_{\sigma}^{\phantom \dagger}(\tau_2) \rangle \langle T_{\tau} c_{\sigma'}^{\dagger}(\tau_3) 
c_{\sigma'}^{\phantom \dagger}(0) \rangle \nonumber] \ .
\end{eqnarray}
Equation~(\ref{equ:gamma}) can be used to rewrite Eq.~(3) of the main text in the following way:
\begin{equation}
\label{equ:BSE-chis}
  \chi_{\mathbf{q}}(\Omega)=\frac{1}{\beta^2}\sum_{\nu \nu'}\left[[\chi^{\nu\nu'}\!(\Omega)]^{-1}+ [\chi^0_{\mathbf q}(\Omega)]^{-1}_{\nu \nu'}-[\chi^{\nu\nu'}_0(\Omega)]^{-1}\right]^{-1}\mkern-18mu,
\end{equation}
where the bubble term reads explicitly
\begin{equation}
  \chi^{0,\nu \nu'}_{\mathbf q}(\Omega) = \frac{-2\beta}{V}\sum_{\mathbf k} \frac{1}{\zeta_\nu-\epsilon_{\mathbf k}} \frac{1}{\zeta_{\nu+\Omega}-\epsilon_{{\mathbf k}+{\mathbf q}}} \delta_{\nu \nu'} \, ,
\end{equation}
with $\zeta_\nu=i\nu+\mu-\Sigma(\nu)$ and $\epsilon_{\mathbf k}=-\frac{2t}{\sqrt{2d}} \sum_i^d \cos k_i$. We can reformulate the bubble terms with the Hilbert transform defined as:
\begin{equation}
  \mathcal{H}(\zeta)=\int^{+\infty}_{-\infty} \!\mathrm{d} \epsilon\, D(\epsilon)\frac{1}{\zeta-\epsilon} \, ,
\end{equation}
where $D(\epsilon)$ is the non-interacting density of states. By summing over all momenta $\mathbf{q}$ we obtain the local bubble term. Thereby, the sums over the two different momenta factorize and we get 
\begin{equation}
  \chi_0^{\nu \nu'}(\Omega) = \frac{1}{V}\sum_{\mathbf{q}}\chi^{0,\nu\nu'}_{\mathbf{q}}(\Omega) = -2\beta\mathcal{H}(\zeta_\nu)\mathcal{H}(\zeta_{\nu+\Omega})\delta_{\nu \nu'}\, .
\end{equation}
In our case of a static charge response ($\mathbf{q}\rightarrow 0,\, \Omega=0 $) the $\mathbf{q}$-dependent bubble reads
\begin{equation}
\begin{split}
  \chi^{0,\nu\nu'}_{\mathbf{q}=0}(\Omega=0) &= - 2\beta\int^{+\infty}_{-\infty}\! \mathrm{d} \epsilon\, D(\epsilon)\frac{1}{(\zeta-\epsilon)^2}\delta_{\nu \nu'}\\
  &= 2\beta \frac{\mathrm{d}\mathcal{H}(\zeta_\nu)}{\mathrm{d}\zeta_\nu}\delta_{\nu \nu'}.
\end{split}
\end{equation}

\subsection{Bethe-lattice}
\label{sec:11}
For the Bethe-lattice with a semi-elliptic density of states the Hilbert transform simplifies to
\begin{equation}
  \mathcal{H}(\zeta)=\frac{\zeta-\mathrm{sgn}(\mathrm{Im}\zeta)\sqrt{\zeta^2-4t^2}}{2t^2}\, ,
\end{equation}
and the difference between the two inverted bubble terms is equal to a constant:
\begin{equation}
\label{equ:bubbleterms}
[\chi^0_{\mathbf{q}=0}]_{\nu\nu'}^{-1} - [\chi_0^{\nu\nu'}]^{-1} = \frac{t^2}{2\beta}\delta_{\nu \nu'}\, .
\end{equation}
We note that this result does not depend on the frequency $\nu$ or on the filling (or the chemical potential $\mu$ whose information is encoded in $\zeta$).

By going into the eigenbasis of the generalized susceptibility
\begin{equation}
  \frac{1}{\beta^2}\sum_{\nu \nu'} \chi^{\nu \nu'}(\Omega=0)=\sum_\alpha \lambda_\alpha w_\alpha \, ,
\end{equation}
we obtain Eq.~(4) of the main text
\begin{equation}
  \kappa = \chi_{\mathbf{q}=0}(\Omega=0) = \sum_{\alpha}\Big (\frac{1}{\lambda_\alpha}+\beta t^2/2\Big)^{-1} w_\alpha \, .
\end{equation}

By a similar derivation but inserting the particle-particle DMFT bubble at $\mathbf{q}=0$ or the particle-hole bubble at $\mathbf{q}=(\pi,\pi,\pi,...)$ into Eq.~(\ref{equ:bubbleterms}) one can show that one gets an additional minus sign for $\frac{t^2}{2\beta}$. 

\subsection{Square lattice}

As discussed in the main text, the results for the one-band Hubbard model on a square lattice 
can be readily understood using Eq.~(4) of the main text, where $t^2\rightarrow t_{\rm eff}^2$.
In this section we discuss the validity of this approach for our parameter set. 

\begin{figure}[h!]
  \centering
  {{\resizebox{8.5cm}{!}{\includegraphics {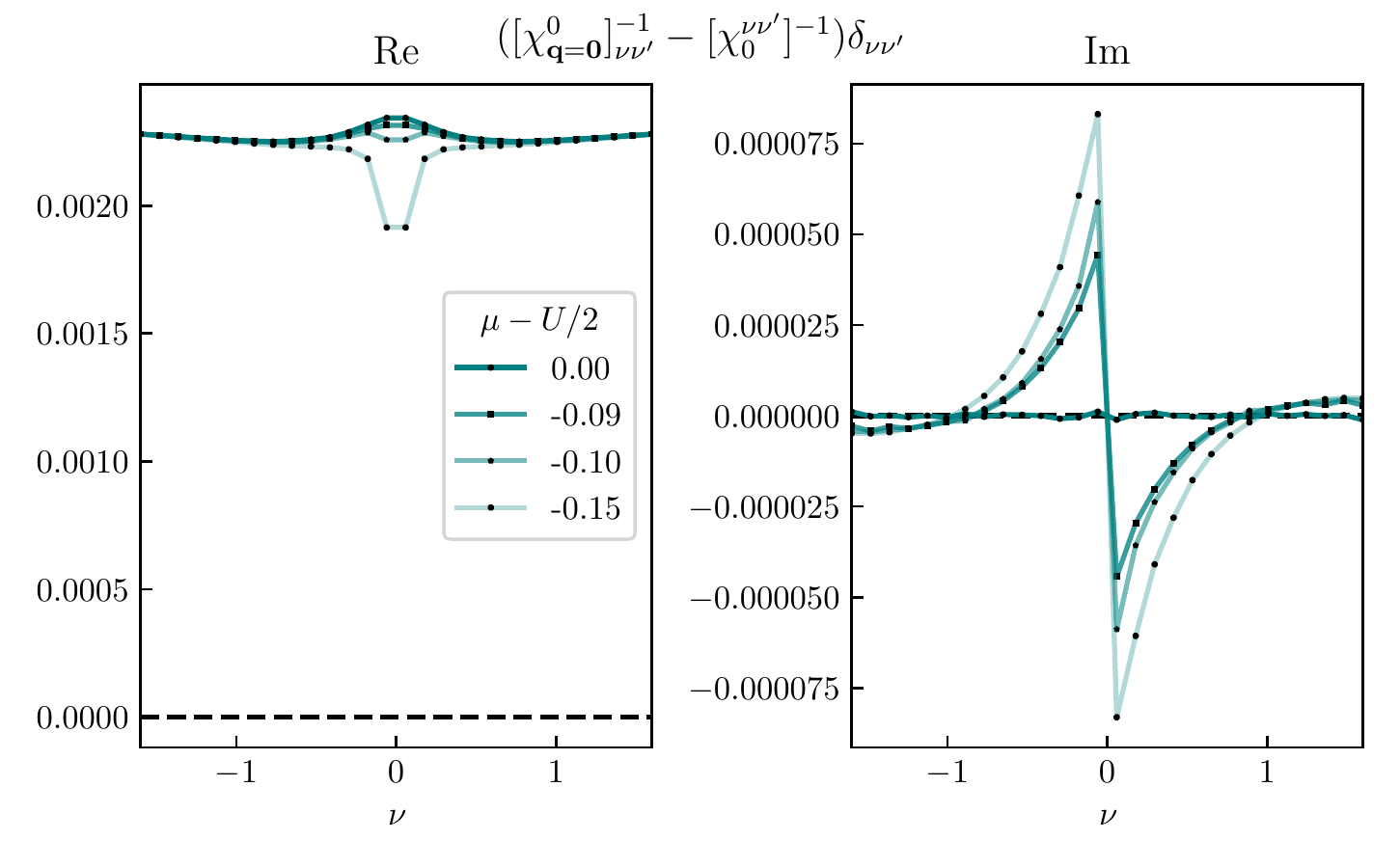}}} 
  \caption{{\small The difference between the inverse of the static limit of the momentum dependent and the local bubble term ($[\chi^0_{\mathbf{q}=0}]_{\nu\nu'}^{-1}-[\chi_0^{\nu\nu'}]^{-1}$) at different chemical potential ($\mu-U/2$) and as function of fermionic Matsubara frequency ($\nu$), showing an almost constant behavior. Left: real part. Right: imaginary part.}} 
  \label{fig:Fig1}  
  }
  \end{figure}

\begin{figure}[h!]
\centering
{{\resizebox{8.5cm}{!}{\includegraphics {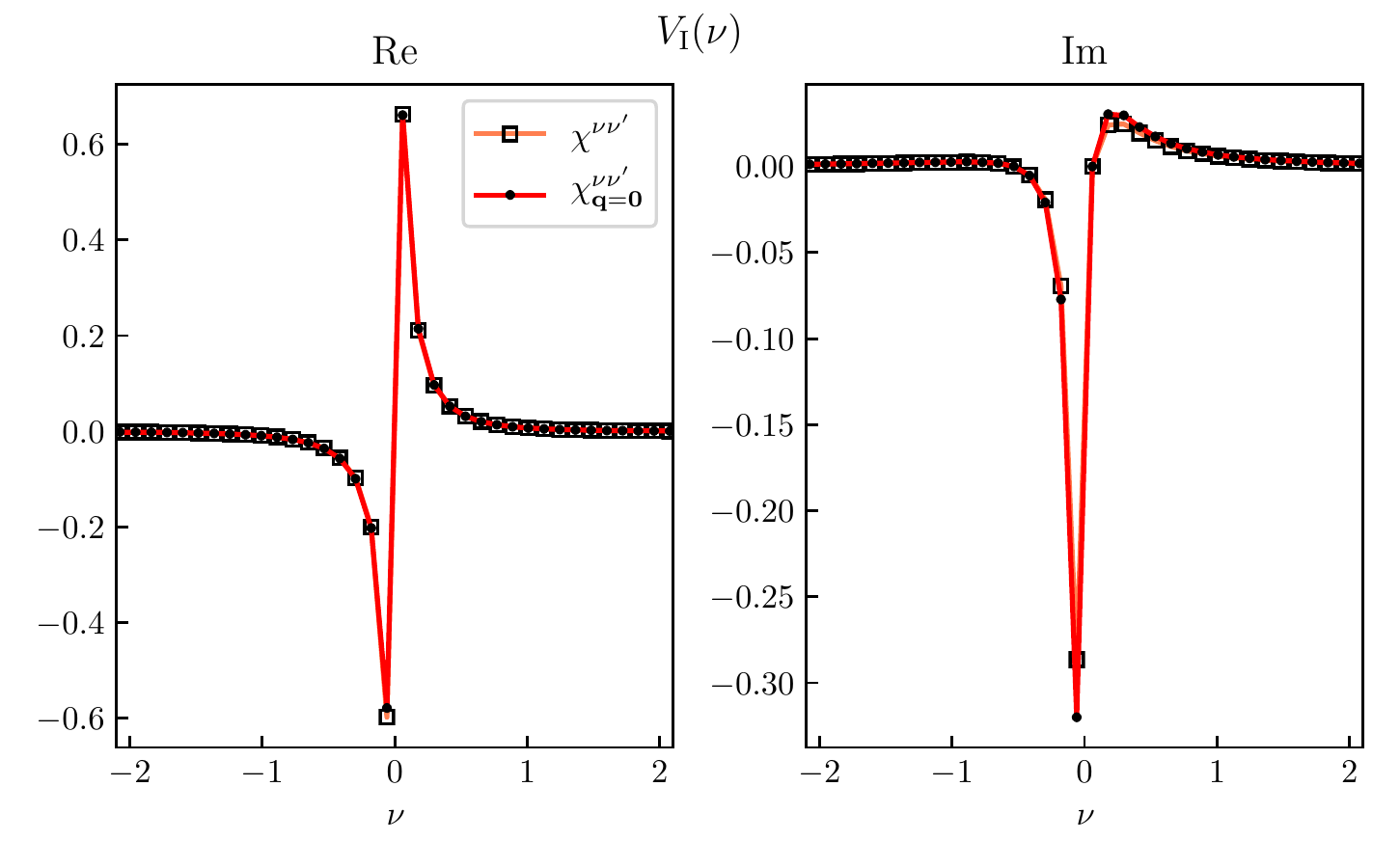}}} 
\caption{{\small Comparison of the eigenvectors $V_\RN{1}(\nu)$ of the lowest real eigenvalue of $\chi^{\nu\nu'}$ and $\chi_{\mathbf{q=0}}^{\nu\nu'}$ at the maximum of $\kappa$ ($U=2.4$, $\beta=53$, $\mu-U/2=-0.1$). Left: real part. Right: imaginary part.
}}
\label{fig:Fig2}
}
\end{figure}

Above we showed, that the term $[\chi^0_{\mathbf{q}=0}]_{\nu\nu'}^{-1} - [\chi_0^{\nu\nu'}]^{-1}$
is equal to the constant $\frac{t^2}{2\beta} \delta_{\nu\nu'}$ for the Bethe-lattice. Using the approximation $t^2\rightarrow t_{\rm eff}^2$
hence boils down to assuming that the difference of the $\mathbf{q}$-dependent and local inverted bubble terms 
for the square lattice is 
constant in Matsubara frequency space. Furthermore, other than for the Bethe-lattice case, the difference of the inverted bubble terms is in principle also depending on the filling. The behaviour of $[\chi^0_{\mathbf{q}=0}]_{\nu\nu'}^{-1} - [\chi_0^{\nu\nu'}]^{-1}$ is explicitly shown in Fig.~\ref{fig:Fig1} (we recall that 
this quantity only depends on one 
Matsubara frequency $\nu$ since both matrices are diagonal in the fermionic frequency 
space). The result of this calculations are summarized in the upper panel of Fig.~3 of the main text, where 
we show, for each $\mu$, the variation of the real part of $[\chi^0_{\mathbf{q}=0}]_{\nu\nu'}^{-1} - [\chi_0^{\nu\nu'}]^{-1}$ as a function of Matsubara frequency $\nu$ (blue-shaded area). It can be clearly seen that this variation is small with respect to the difference of the two lowest real eigenvalues of $\chi^{\nu \nu'}(\Omega=0)$ (which are central to our study). This allows us to restrict the analysis of the phase-separation instability in DMFT to the lowest eigenvalue $\lambda_{\RN{1}}$. In the right panel of Fig.~\ref{fig:Fig1} we also see that the imaginary part (zero in the Bethe 
lattice case) is nonzero but vanishingly small and therefore negligible. Hence using Eq.~(4) of the main text as a basis for 
the explanation is a valid approach. 
This statement can be further strengthened by comparing the eigenvectors $V_{\alpha}(\nu)$ of the local 
$\chi^{\nu\nu'}(\Omega=0)$ and the uniform generalized charge susceptibility $\chi_{\mathbf{q=0}}^{\nu\nu'}(\Omega=0)$, where $\kappa = \frac{1}{\beta^2}\sum_{\nu\nu'}\chi_{\mathbf{q=0}}^{\nu\nu'}(\Omega=0)$. The comparison is made for the parameters corresponding to the maximum of $\kappa$ and focusing on the eigenvector corresponding to $\lambda_{\rm I}$, i.e. the eigenvector associated to the most negative eigenvalue of $\chi_{\mathbf{q=0}}^{\nu\nu'}(\Omega=0)$ (as 
discussed in the main text). 
A perfect agreement of the two eigenvectors is obviously found only in the Bethe-lattice case, 
but as it is shown in Fig.~\ref{fig:Fig2}, also for the square lattice the agreement is very convincing.\\

In this context, where one restricts the analysis to the lowest eigenvalues of $\chi^{\nu\nu'}(\Omega=0)$, it is possible to provide a precise definition of $\beta t^2_{\rm eff}/2$: for each value of $\mu$, we determine the value of $\widetilde{\lambda_{\rm I}}$, that would trigger $\kappa (\widetilde{\lambda}_\RN{1} \in \mathbb{R}) \rightarrow \infty$ by using the BSE (Eq.~(\ref{equ:BSE-chis})). Then the value of $t_{\rm eff}$ is determined from: $\frac{1}{\widetilde{\lambda}_\RN{1}}-\beta t^2_{\rm eff}/2 = 0$.

This definition is used to mark the blue dashed line in the upper panel of Fig.~3 of the main text. Evidently the physical maximum of $\kappa$ corresponds to the minimal difference between $\lambda_\RN{1}$ and the value of $-2/(\beta t^2_{\rm eff})$. 
Since $\lambda_{\rm I}$ never reaches this condition, we confirm our statement in the main text that for $\beta=53$, $U=2.4$ we are just slightly above the onset of the phase separation. It is interesting to note that the value 
$\beta t^2_{\rm eff}/2$ corresponds, with a satisfying level of agreement to the value of $[\chi^0_{\mathbf{q}=0}]_{\nu\nu'}^{-1} - [\chi_0^{\nu\nu'}]^{-1}$ for the lowest Matsubara frequency. This is consistent with the observation that the eigenvector $V_{\RN{1}}(\nu)$ of $\lambda_{\RN{1}}$ (see Fig.~\ref{fig:Fig3} and \cite{Chalupa2018}) is extremely localized in the low-frequency domain.

As both, the overall frequency dependence and the $\mu$-dependence are weak compared to $\mid\lambda_{\RN{1}}-\lambda_{\RN{2}}\mid$ the approximation based on the Bethe-lattice expression works reasonably well for the square lattice. As a result, the fulfilment of the condition for the enhancement/divergence of $\kappa$ matches to a good approximation 
the minimum value of $\lambda_{\RN{1}}$.

\section{II. TECHNICAL DETAILS OF THE DMFT CALULATIONS}

Data throughout the main text and this supplemental material were obtained from a continuous-time quantum Monte Carlo impurity solver in the hybridization expansion (CT-HYB), as implemented in the w2dynamics package (version 1.0.0)\cite{w2dynamics}.

The computations on the one-particle level for the Bethe-lattice were performed at half-filling, $\mu=U/2$. In the square lattice case the different DMFT data-points were obtained via $\mu$-fixed iterations using $48 \times 48$ k-points.

Prior to the direct calculation of the generalized susceptibilities defined in Eq.~(\ref{equ:form_gen_chi}), convergence of the DMFT algorithm for the one-particle quantities was achieved. By carefully converging from multiple starting points, we ruled out coexisting solutions at the maximum of $\kappa$. 
For the final calculations of the two-particle quantities with $260 \times 260$ fermionic Matsubara frequencies we have used up to 1.600 CPU cores; on each core we performed $\mathcal{O}(10^7)$
warm-up sweeps and $\mathcal{O}(10^8)$ Monte Carlo simulation steps, during which we have been measuring every 60 steps in order to reduce autocorrelation.  
In frequency summations the missing high frequency contributions have been approximated by the $\mathcal{O}(1/\nu^2)$-dependence of the bubble-term up to infinite frequencies.

Finally for the calculation of the uniform susceptibility $\chi_{\mathbf{q}}^{\nu\nu'}(\Omega=0)$ the code available at 
[\onlinecite{ladderDGA}] was employed where 100 k-points were used in the internal summations of the Bethe-Salpeter equation and 55 $\mathbf{q}$-points for Fig.~4 in the main text. The numerical derivative $\frac{d n}{d \mu}$ in Fig.~2 was calculated via central finite differences. Right at the maximum of $\kappa$, in the close proximity of the critical endpoint, the convergence of the DMFT calculation requires an increasingly large number of iterations. For our calculations these effects lead to an intrinsic uncertainty of the maximum value of the order of $\Delta \kappa \simeq 10 \%$.

\section{III. Spectral properties}

\subsection{Correspondence with vertex divergences}

\begin{figure}[h!]
  \centering
  {{\resizebox{8.5cm}{!}{\includegraphics {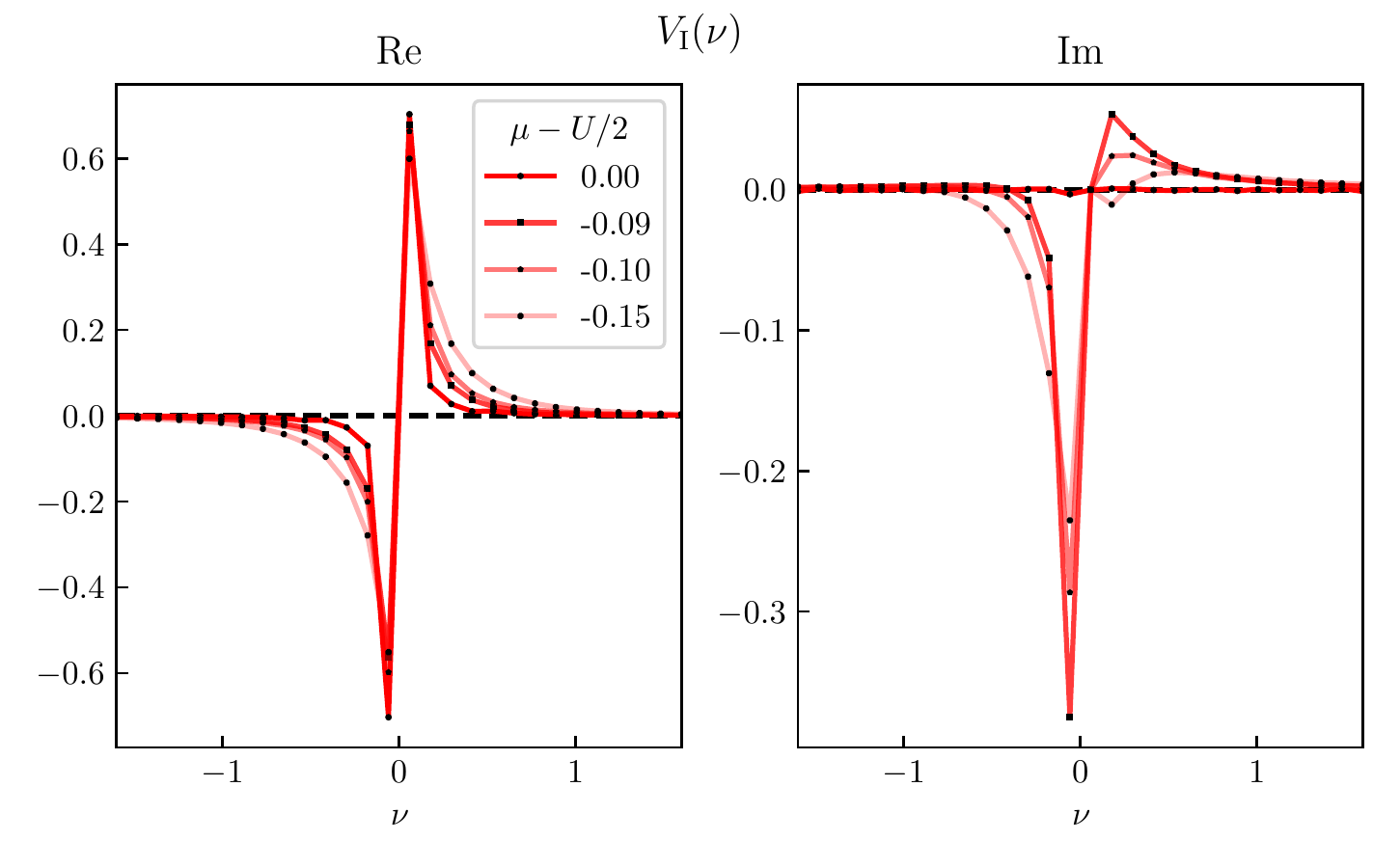}}}
  {{\resizebox{8.5cm}{!}{\includegraphics {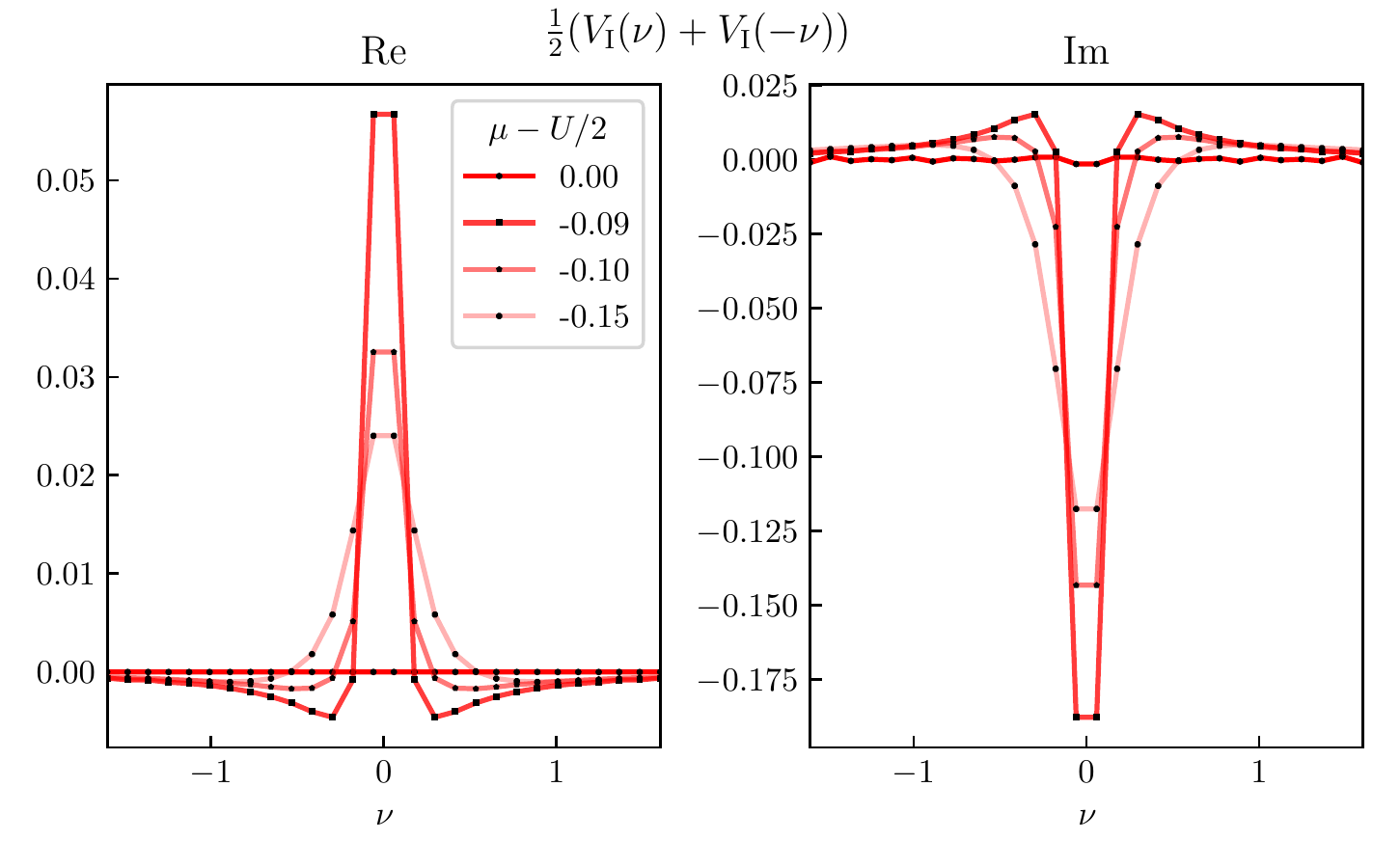}}}} 
  \caption{{\small Upper: Gradual evolution of the eigenvector $V_\RN{1}(\nu)$, corresponding to $\lambda_\RN{1}$, from half-filling ($\mu-U/2=0$) to finite doping ($\mu-U/2=-0.15$). Lower: Evolution of the symmetrized eigenvector $\frac{1}{2}(V_\RN{1}(\nu)+V_\RN{1}(-\nu))$, corresponding to $\lambda_\RN{1}$, highlighting the antisymmetry of $V_\RN{1}(\nu)$ at half-filling. At finite doping, the condition $V_\RN{1}(\nu)=-V_\RN{1}(-\nu)$ is violated, and the symmetrized eigenvector shows non-zero values. Left: corresponding real parts. Right: corresponding imaginary parts.}} 
    \label{fig:Fig3}
  }
\end{figure}

\begin{figure}[h!]
  \centering
  {{\resizebox{8.5cm}{!}{\includegraphics {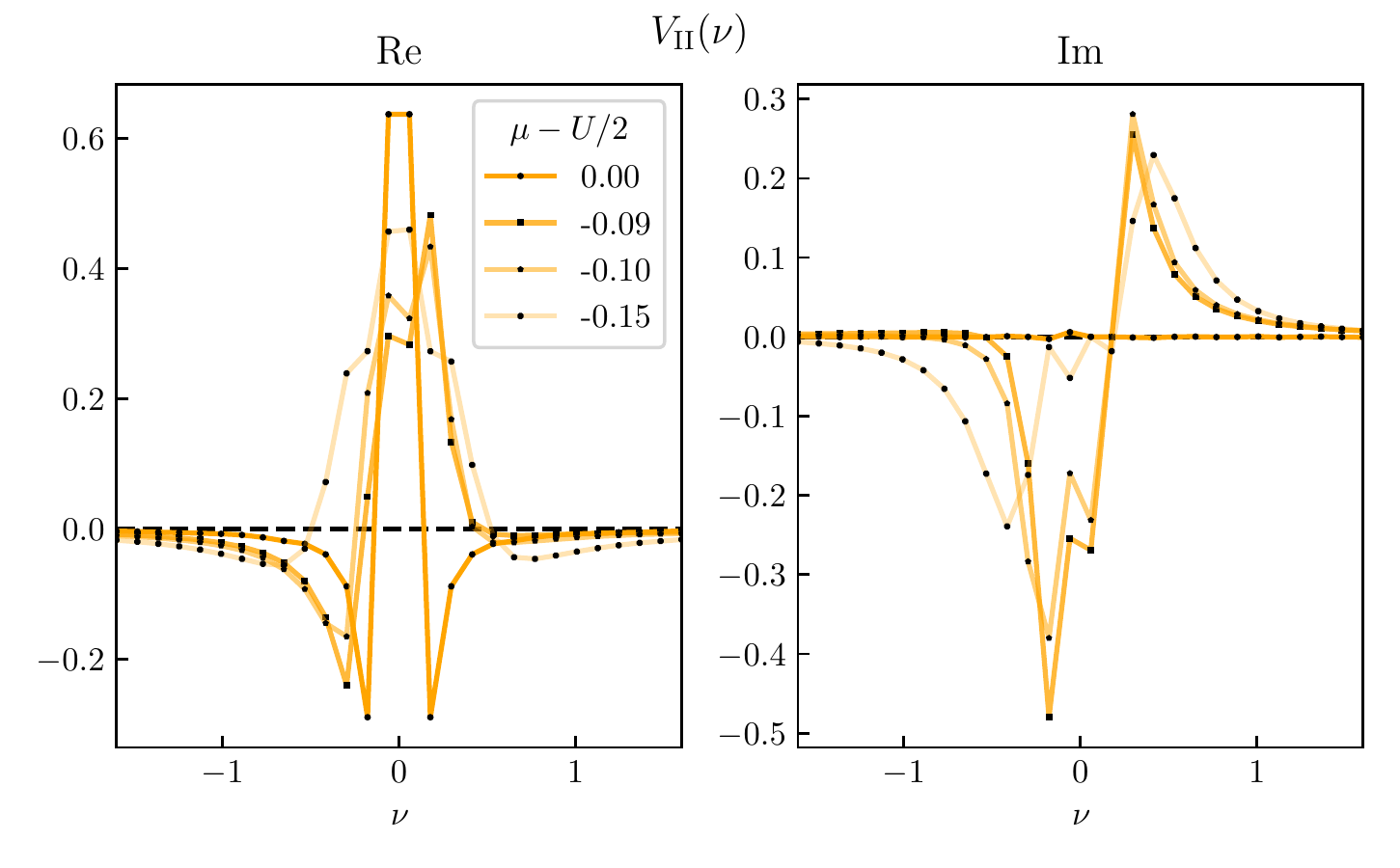}}} 
  \caption{{\small Gradual evolution of the eigenvector $V_\RN{2}(\nu)$, corresponding to $\lambda_\RN{2}$, from half-filling ($\mu-U/2=0$) to finite doping ($\mu-U/2=-0.15$). Left: real part. Right: imaginary part.}} 
  \label{fig:Fig4}
  }
\end{figure}

In this section we show, that the eigenvalues $\lambda_{\rm I}$ and $\lambda_{\rm II}$, discussed in the main text, 
are directly related to the appearance of the first and second vertex divergence lines 
(line I and II in Fig.~1a of the main text).
To this end, we study the continuous evolution of the eigenvectors $V_{\rm I/II}(\nu)$ 
(corresponding to the eigenvalues $\lambda_{\rm I/II}$) of the local charge susceptibility $\chi^{\nu \nu'}(\Omega=0)$
as a function of filling, see Fig.~\ref{fig:Fig3} and \ref{fig:Fig4}. 
As stated in the main text, at half-filling, i.e. $\mu-U/2=0.0$, $\chi^{\nu \nu'}(\Omega=0)$ is a real, bisymmetric matrix. This implies\cite{Springer2020}, that the eigenvectors $V_{\rm I/II}(\nu)$ are real and either symmetric, or 
antisymmetric, with respect to $\nu\leftrightarrow-\nu$.
Fig.~\ref{fig:Fig3} shows readily that the real part of $V_{\rm I}(\nu)$ is antisymmetric at half-filling (left panel),
 whereas the 
imaginary part is vanishing (right panel). The same holds for $V_{\rm II}(\nu)$ (see Fig.~\ref{fig:Fig4}), whereas it is symmetric 
at half-filling. 
As discussed in several works\cite{Chalupa2018,Thunstroem2018,Springer2020} on the appearance of vertex divergences in 
fundamental models of many-electrons systems, the first divergence line is associated with an antisymmetric eigenvector, 
the second line with a symmetric one. 
For high- and intermediate temperatures\cite{Schaefer2016c,Chalupa2018, Springer2020} the eigenvector of the 
first divergence line resembles the one of the atomic limit\cite{Schaefer2016c,Thunstroem2018} 
($\frac{1}{\sqrt{2}}(\delta_{\nu\bar{\nu}} - \delta_{\nu\bar{\nu}})$), where $\bar{\nu}=\pi T$. This is clearly 
recognizable in Fig.~\ref{fig:Fig3}. On the other hand the eigenvector of the second divergence line\cite{Chalupa2018,Springer2020} 
has similarities with the symmetric combination ($\frac{1}{\sqrt{2}}(\delta_{\nu\bar{\nu}} + \delta_{\nu\bar{\nu}})$), 
also apparent in Fig.~\ref{fig:Fig4}.
Hence, the eigenvalues $\lambda_{\rm I}$ and $\lambda_{\rm II}$ and their corresponding eigenvectors $V_{\rm I}(\nu)$ and $V_{\rm I}(\nu)$ are those that also originate the divergence lines I and II at lower values of $U$.

\subsection{Negative weights}

As already discussed in earlier literature\cite{Rohringer2012, Thunstroem2018,Springer2020} at half-filling, when the system fulfills particle-hole symmetry -together with time-reversal and SU$(2)$-symmetry- $\chi^{\nu \nu'}(\Omega\!=\!0)$ obeys the conditions 
\begin{equation}
  \chi^{\nu \nu'}=(\chi^{\nu \nu'})^*=\chi^{(-\nu) (-\nu')} \quad \mathrm{and} \quad \chi^{\nu \nu'}\!=\chi^{\nu'\!\nu},
\end{equation}
and is therefore a real and {\sl bisymmetric} matrix. However, at finite doping the particle-hole symmetry of the system is violated.
As a consequence, due to remaining time-reversal and SU$(2)$-symmetry, the following conditions for $\chi^{\nu \nu'}(\Omega\!=\!0)$ hold:
\begin{equation}
  (\chi^{\nu \nu'})^*=\chi^{(-\nu) (-\nu')} \quad \mathrm{and} \quad \chi^{\nu \nu'}\!=\chi^{\nu'\!\nu}.
\end{equation}
Hence $\chi^{\nu \nu'}(\Omega\!=\!0)$ 
is no longer real or even hermitian, but {\sl centrohermitian}\cite{centroherm} (and symmetric). This enables the possibility of negative weights $w_\alpha <0$ and complex conjugate pairs of $\lambda_\alpha$ for the out of half-filling case, as discussed in
the main text.

For $\lambda_{\rm I}$ ($\in \mathbb{R}$) the negative weight
\begin{equation}
  \begin{split}
w_{\rm I} \! &=\!  {\rm Re} [\sum_\nu V^{-1}_{\rm I}(\nu)]  \, {\rm Re} [\sum_{\nu'} V^{}_{\rm I}(\nu')] \\
&- {\rm Im}[\sum_\nu V^{-1}_{\rm I}(\nu)] \, {\rm Im}[\sum_{\nu'} V^{}_{\rm I}(\nu')] 
  \end{split}
\end{equation}
is originated from {\sl both} summands, whereas the one stemming from the imaginary part is found to be the dominant one. Note that, due to the centrohermitian properties of $\chi^{\nu \nu'}(\Omega=0)$, all $w_\alpha$ are real.

\section{IV. Comparison with half-filling}

\begin{figure}[h!]
  \centering
  {{\resizebox{8.5cm}{!}{\includegraphics {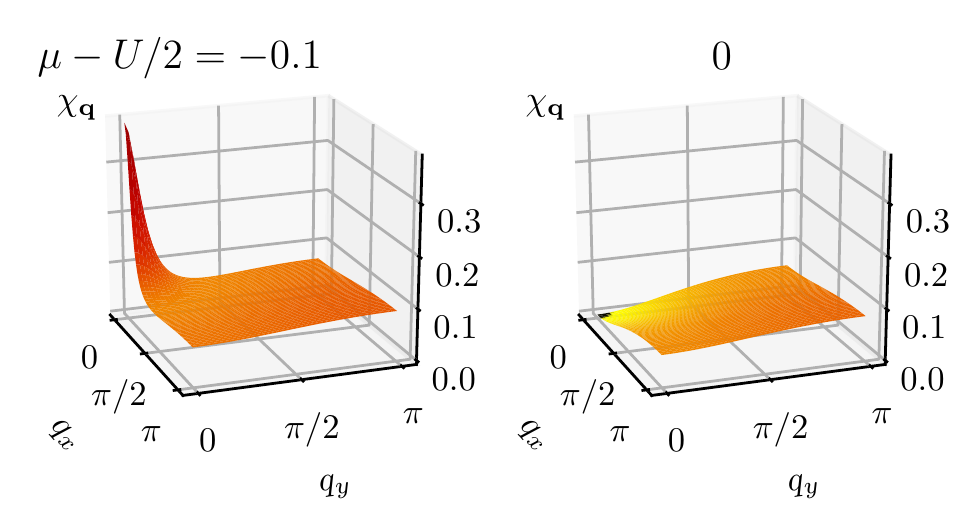}}} 
  {{\resizebox{8.5cm}{!}{\includegraphics {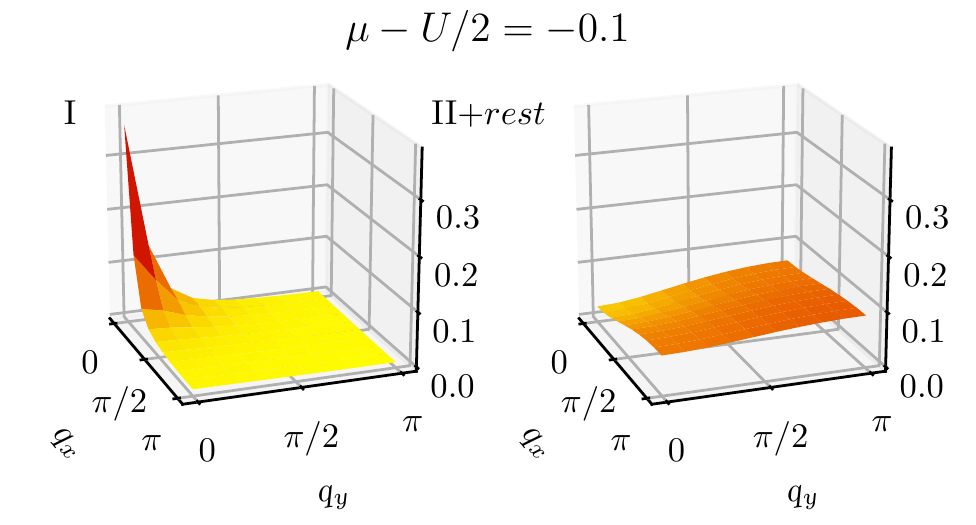}}}}
  \caption{{\small Upper: Comparison of the momentum dependence of $\chi_{\mathbf{q}}$ for the maximum of $\kappa$ 
  ($\mu-U/2=-0.1$) and half-filling ($0$). 
  Lower: The ($\mu-U/2=-0.1$) result separated into the critical attractive contribution 
  originated from $\lambda_{\rm I}$ (left) and the rest (right).}} 
  \label{fig:Fig5}
  }
\end{figure}

In the main text we discuss that without the critical attractive contribution stemming from $\lambda_{\rm I}$, 
$\chi_{\mathbf{q}}(\Omega=0)$ at the maximum of $\kappa$, i.e. for $\mu-U/2=-0.1$, closely resembles the one at half-filling. 
This is shown in Fig.~\ref{fig:Fig5}, where for the sake of completeness the 
result for $\chi_{\mathbf{q}}(\Omega=0)$ for $\mu-U/2=-0.1$
is repeated. 

\bibliography{VERTEX-SM}


\end{document}